\documentclass[aps,twocolumn,showpacs,preprintnumbers,prb,superscriptaddress]{revtex4}
\usepackage{graphicx}
\usepackage{amssymb}
\usepackage[toc,page]{appendix}
\newcommand{\R}{\mathbf{r}}

\newcommand{\be}{\begin{equation}}
\newcommand{\ee}{\end{equation}}
\newcommand{\bea}{\begin{eqnarray}}
\newcommand{\eea}{\end{eqnarray}}
\newcommand{\bean}{\begin{eqnarray*}}
\newcommand{\eean}{\end{eqnarray*}}

\begin{document}

\title{Semilocal Density Functional Theory with correct surface asymptotics}
\author{Lucian A. Constantin}
\affiliation{Center for Biomolecular Nanotechnologies @UNILE, Istituto Italiano di Tecnologia, Via Barsanti, I-73010 Arnesano, Italy}
\author{Eduardo Fabiano}
\affiliation{Istituto Nanoscienze-CNR, Euromediterranean Center for Nanomaterial Modelling and Technology (ECMT), via Arnesano 73100, Lecce}
\affiliation{Center for Biomolecular Nanotechnologies @UNILE, Istituto Italiano di Tecnologia, Via Barsanti, I-73010 Arnesano, Italy}
\author{J. M. Pitarke}
\affiliation{CIC nanoGUNE, Tolosa Hiribidea 76, E-20018 Donostia, Basque Country}
\affiliation{Materia Kondentsatuaren Fisika Saila, DIPC, and Centro F\'\i sica Materiales CSIC-UPV/EHU,\\
644 Posta kutxatila, E-48080 Bilbo, Basque Country}
\author{Fabio Della Sala}
\affiliation{Istituto Nanoscienze-CNR, Euromediterranean Center for Nanomaterial Modelling and Technology (ECMT), via Arnesano 73100, Lecce}
\affiliation{Center for Biomolecular Nanotechnologies @UNILE, Istituto Italiano di Tecnologia, Via Barsanti, I-73010 Arnesano, Italy}

\date{\today}

\begin{abstract}
Semilocal Density Functional Theory
is the most used computational method for
electronic structure calculations in theoretical solid-state physics
and quantum chemistry of large systems, providing good accuracy with a very attractive
computational cost. Nevertheless, because of the non-locality of the
exchange-correlation hole outside a metal surface, it was always considered inappropriate to describe the
correct surface asymptotics.
Here, we derive, within the semilocal Density Functional Theory formalism,
an exact condition for the image-like surface
asymptotics of both the exchange-correlation energy per particle
and potential.
We show that this condition can be easily incorporated into a practical
computational tool,
at the simple
meta-generalized-gradient approximation level of theory.
Using this tool, we also show that the Airy-gas model exhibits asymptotic
properties that are closely related to the ones at metal surfaces.
This result highlights the relevance of the linear effective potential model to the
metal surface asymptotics.
\end{abstract}


\maketitle

\section{Introduction}

The exact form of the potential felt by an electron leaving from or
approaching a metal surface is of great importance for a variety of
physical phenomena, including the interpretation of image states \cite{garcia85}, 
modeling quantum-transport \cite{datta2005quantum},
low-energy electron diffraction (LEED)~\cite{RM}, 
scanning tunneling microscopy~\cite{Binning85,Pitarke90}, 
and inverse or two-photon photoemission 
spectroscopy \cite{harris97,fauster}.
The asymptotic form of this image potential is 
$-1/(4(z-z_0))$, with $z$ being the distance from the surface, and $z_0$ representing the position of 
the so-called image plane\cite{langPRB73}, and should be reproduced by any 
computational method aiming at an accurate description
of the surface physics.

Within Kohn-Sham (KS) density-functional theory (DFT)~\cite{kohn99,kohn1965self}, 
which is the most used computational method for
electronic structure calculations in theoretical solid-state physics,
the shape of the image potential is dictated by the properties
of the effective Kohn-Sham (KS) potential. This 
depends on the employed approximation for the
exchange-correlation (XC) functional $E_{xc}[\rho]$, which
gives the XC potential 
via the relation
\begin{equation}
v_{xc}({\bf r})={\delta E_{xc}[\rho]\over \delta \rho({\bf r})},
\end{equation}
where $\rho$ is the electron density.
It has been shown that the exact $v_{xc}$ asymptotically approaches the image potential \cite{langPRB73,gunn79,eghanke89,egu92,white98}, despite
a different result has been obtained within the plasmon-pole approximation\cite{qian2005exact}.

The popular local density approximation (LDA) \cite{kohn1965self} and 
the generalized-gradient approximation (GGA), however,
fail in this task \cite{hoeft2001molecular} showing either
a too fast decay (e.g. exponential), or an
inaccurate description of the surface energetics
(as for the Becke exchange \cite{b88}).
Ad-hoc modifications of the LDA XC potential \cite{serena86,Chulk99}
have then been proposed to improve the asymptotic behavior of the XC potential,
but such methods are not functional derivatives of 
any energy functional. Alternatively, 
non-local methods outside the KS
framework \cite{gunn79,gunn80,ossicini86,garcia00,egu92,white98}
are employed.

An accurate KS-DFT method 
with the correct surface asymptotics would 
be desirable for many reasons, including
the local nature and the computational efficiency.
However, a good functional shall yield not only
the correct asymptotic XC potential, but also accurate energies.
Thus, it is necessary 
to be defined by a realistic
energy per particle $\varepsilon_{xc}({\bf r})$. The latter
is not a uniquely defined physical quantity, but an exact
reference for it is the the conventional
energy per particle, which is associated with the interaction of 
an electron at ${\bf r}$ with the coupling-constant-averaged 
charge of its XC hole~\cite{HG,LP,GL}.
The exact (conventional) $\varepsilon_{xc}({\bf r})$ at metal 
surfaces decays as  $-1/(4(z-z_0))$, i.e. 
as the image potential \cite{gunn79,CP2}.

We recall that the exact exchange energy per particle decreases as $\epsilon_x(z\rightarrow\infty)\rightarrow -A(\beta)/z$ where 
$A(\beta)=(\pi+2\beta \ln(\beta))/(2\pi(1+\beta^2))$, $\beta=\sqrt{\epsilon_F/W}$, $\epsilon_F=k_F^2/2$ being the Fermi energy 
(and $k_F$ the Fermi wavevector), and 
$W$ the work function \cite{CP3,qian2012asymptotic}. 
On the other hand, the exact exchange potential behaves as 
$v_x(z\rightarrow\infty)\rightarrow \ln(\beta k_Fz)/(2\pi\beta z)$\cite{HPR2}. 
Note that these behaviors are related to semi-infinite surfaces; for finite jellium slabs  
we have that, as in molecules, $\epsilon_x(z\rightarrow\infty)=-1/(2 z)$ \cite{CP3} and   
$v_x(z\rightarrow\infty)=-1/(z)$\cite{horo06,horo08,ye15,engel2014exact,engel2014asymptotic}.  

The simultaneous description of surface asymptotic and energy properties 
is anyway an ambitious objective, which is in fact
not achievable at the GGA level \cite{engel92,lb94,arm13,alpha}.
In this article, we show that the issue can be instead solved 
at the meta-GGA level of theory, employing 
an exact condition which yields the correct
image-like asymptotic behavior of both $\varepsilon_{xc}$ and $v_{xc}$ at 
metal surfaces. This condition can be easily 
implemented in any meta-GGA functional, keeping its original accuracy 
for ground-state properties not related to surface asymptotics. 
Hence, an accurate KS-DFT method with correct metal-surface
asymptotic can be obtained for application in many surface science problems.

\section{Exact condition for asymptotic properties}

To start, we consider the simplest (and most used) model for a metal 
surface: the semi-infinite jellium surface. 
This model system is very important in surface science and solid-state physics, containing the physics of
simple metal surfaces \cite{LK,langPRB71,langPRB73}.
	
The KS single-particle orbitals have the form
\begin{equation}
\Psi_{k_z,\mathbf{k}_{||}}(\mathbf{r})=\frac{1}{\sqrt{S}\sqrt{L}}e^{i \mathbf{k}_{||}
\mathbf{r}_{||}}\phi_{k_z}(z)\ ,
\label{ej1}
\end{equation}
where $\mathbf{k}_{||}$ and $\mathbf{r}_{||}$ are the 
two-dimensional wave-vector and position vector in the plane $xy$ 
of the surface, $k_z$ and $z$ are the corresponding
components in the direction perpendicular to the surface, 
$S$ and $L$ are the normalization 
area and length, and $\phi_{k_z}(z)$ are the eigenfunctions
of a one-dimensional KS Hamiltonian (for details see Appendix A).

In the vacuum region, far away from the surface
($z\to\infty$), the single-particle orbitals $\phi_{k_z}(z)$ 
behave as \cite{CP3}:
\begin{equation}
\phi_{k_z}(z\rightarrow \infty)\rightarrow \phi_{k_F}(z\rightarrow \infty)e^{-\beta z (k_F-k_z)},
\label{ej3}
\end{equation}
where
\begin{equation}
\phi_{k_F}(z\rightarrow \infty)\sim e^{-z\sqrt{2W}}(2z\sqrt{2W})^{\alpha_{KS}/\sqrt{2W}},
\label{ej4}
\end{equation}
%
$\alpha_{KS}>0$.

Now, we consider a meta-GGA XC energy per particle of the form
\begin{equation}
\epsilon_{xc}^{MGGA}=  \epsilon_{x}^{LDA}   \frac{1}{\eta} 
\frac{8\pi}{3\sqrt{5}}\frac{\sqrt{\alpha}}{\sqrt{\ln(\alpha)}}\ ,
\label{em3}
\end{equation}
where $\eta$ is a parameter to be fixed later, 
$\alpha(\R)=\left[\tau(\R)-\tau^W(\R))/\tau^{TF}(\R)\right]$ is the well-known meta-GGA 
ingredient that measures the non-locality of the kinetic-energy density~\cite{alpha},
with $\tau$, $\tau^W$, and $\tau^{TF}$ being the positive-defined exact KS, von Weizs\"{a}cker, and 
Thomas-Fermi kinetic-energy densities, 
respectively. We recall that
$1/\left[1+\alpha(\R)^2\right]$ is the electron localization function, often used in the 
characterization of chemical bonds \cite{elf}.
Eq. (\ref{em3}) yields the following asymptotic behaviors 
(see the Appendix B for details):
\begin{eqnarray}
\epsilon_{xc}^{MGGA}(z\rightarrow\infty) & \rightarrow & -\frac{1}{\eta}\frac{1}{z}    
+\mathcal{O}(z^{-2}),  \label{eje} \\
       v_{xc}^{MGGA}(z\rightarrow\infty) &\rightarrow &  -\frac{3}{2\eta} 
\frac{1}{z}+\mathcal{O}(z^{-2})\ . \label{ejv}
\end{eqnarray}
Here, the KS potential has been obtained in the generalized KS framework using the
formula~\cite{arb02}
\begin{eqnarray}
v_{xc}(\R)\Psi_i(\R) & = &[\frac{\partial(\rho\epsilon_{xc})}{\partial 
\rho}-\nabla\frac{\partial(\rho\epsilon_{xc})}{\partial \nabla \rho}]\Psi_i
-\frac{1}{2}\nabla(\frac{\partial(\rho\epsilon_{xc})}{\partial\tau})\nabla\Psi_i\nonumber\\
&& -\frac{1}{2}\frac{\partial(\rho\epsilon_{xc})}{\partial\tau}\nabla^2\Psi_i\ .
\label{eint4}
\end{eqnarray}

Equations. (\ref{eje}) and (\ref{ejv}) show 
that, in contrast to previously developed XC functionals, both
$v_{xc}$ and $\epsilon_{xc}$ are proportional to the exact ones:
if $\eta=\eta_1=4$ ($\eta=\eta_2=6$) then the exact energy-density 
(potential) is obtained.
Unfortunately, $\eta_1\neq\eta_2$. Nevertheless, for both values
Eq. (\ref{em3}) yields an asymptotic behavior qualitatively and 
quantitatively significantly beyond the current state-of-the-art. 
We also remark that Eq. (\ref{em3}) is solely based on the properties of the reduced kinetic ingredient $\alpha$. 
However, at the meta-GGA level of theory, other ingredients are also available (e.g. the gradient and the Laplacian of the density)
so that the exact asymptotic description of both $\epsilon_{xc}$ and $v_{xc}$ 
might be achieved.

\section{Practical computational tool}

As a first practical example, 
we consider the case $\eta=\eta_1$ and incorporate the
condition of Eq.~(\ref{em3}) into the popular TPSS meta-GGA
functional \cite{tpss}, using an approach similar to that of 
Ref.~\onlinecite{sll}.
The resulting XC functional will be termed surface-asymptotics (SA) TPSS.
This functional is obtained by simply changing, in the TPSS exchange 
formula, the parameter
$\kappa$ (which determines the asymptotic behavior of the functional)
from its original value of 0.804 to
\begin{equation}
\kappa=\frac{2\pi}{3\sqrt{5}}\frac{\sqrt{\alpha+1}}{\sqrt{a+\ln(\alpha+b)}}\ .
\label{em5}
\end{equation}
The correlation is left unchanged. (Note that the TPSS correlation decays exponentially, 
and thus our XC condition is incorporated in the TPSS exchange functional. This is a common procedure 
for semilocal functionals, which are based on a strong error cancellation between the exchange and 
correlation parts.)
In Eq. (\ref{em5}), the parameters $a=2.413$ and $b=0.348$ have been fixed by 
imposing the constraints: $\kappa=0.804$ for $\alpha=1$ and $\alpha=0$;
whereas, when degenerate orbitals contribute to the tail of the density ($\alpha\rightarrow\infty$),
$\kappa\rightarrow\frac{2\pi}{3\sqrt{5}}\frac{\sqrt{\alpha}}{\sqrt{\ln(\alpha)}}$ (i.e. Eq. (\ref{em3})).
These conditions assure that (i) all the exact constraints satisfied by
the original TPSS exchange functional are preserved and (ii) the new functional
yields the correct image-like asymptotics.
The SA-TPSS functional does not recover locally the Lieb-Oxford 
bound \cite{sll}, as TPSS does, but it satisfies the global 
Lieb-Oxford bound for all known physical systems (e.g. for 
atoms, molecules, solids and surfaces $E_{xc}^{SA-TPSS}\approx E_{xc}^{TPSS}$). 
Moreover, it fulfills locally the simplified version of the Lewin-Lieb bound 
(see Eq. (22) of Ref. \cite{feinblum2014communication}). 

%
\begin{figure}[t]
\includegraphics[width=\columnwidth]{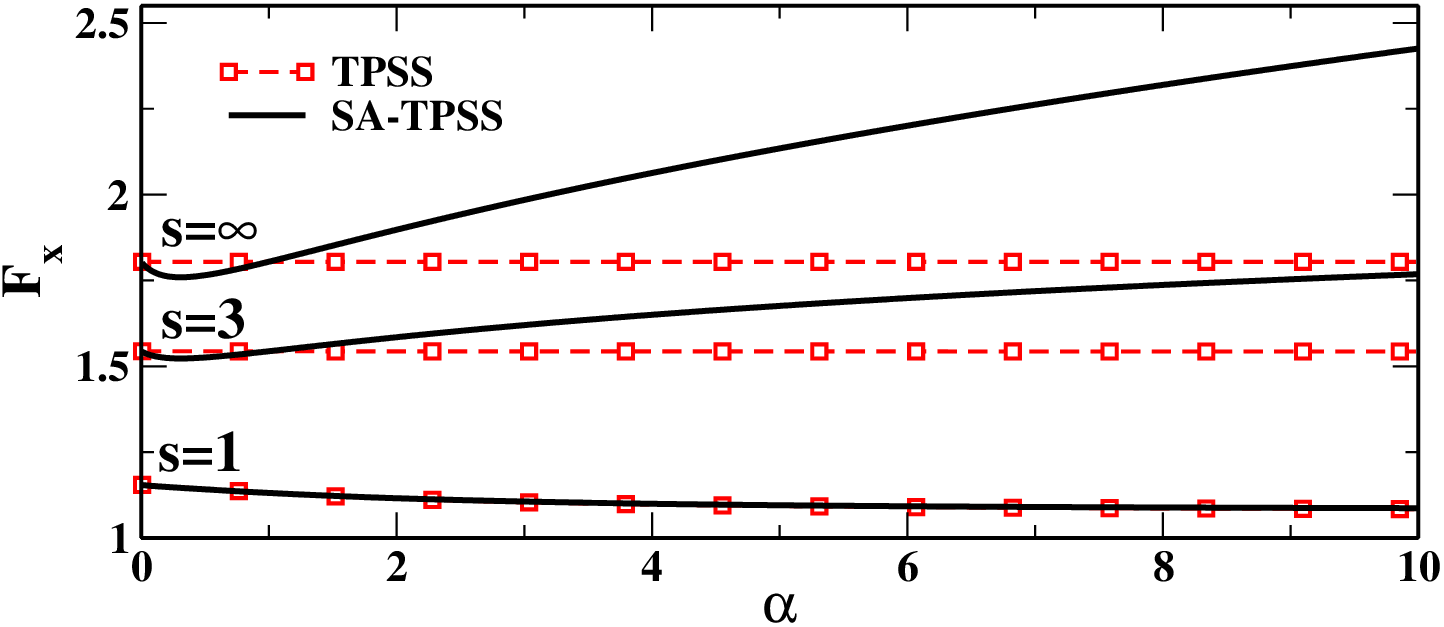}
\caption{TPSS and SA-TPSS exchange enhancement factors versus $\alpha$, 
for three values of the reduced gradient $s$.
}
\label{f1} 
\end{figure} 
%
In Fig. \ref{f1}, we show the TPSS and SA-TPSS exchange 
enhancement factors versus $\alpha$ for several values of the reduced gradient $s=|\nabla \rho|/(2(3\pi^2)^{1/3}\rho^{4/3})$).
When $s$ is small, TPSS and SA-TPSS coincide for all values of $\alpha$. As $s$ increases,
TPSS and SA-TPSS start to differ, especially at large values of $\alpha$, as expected.

%
\begin{figure}[t]
\includegraphics[width=\columnwidth]{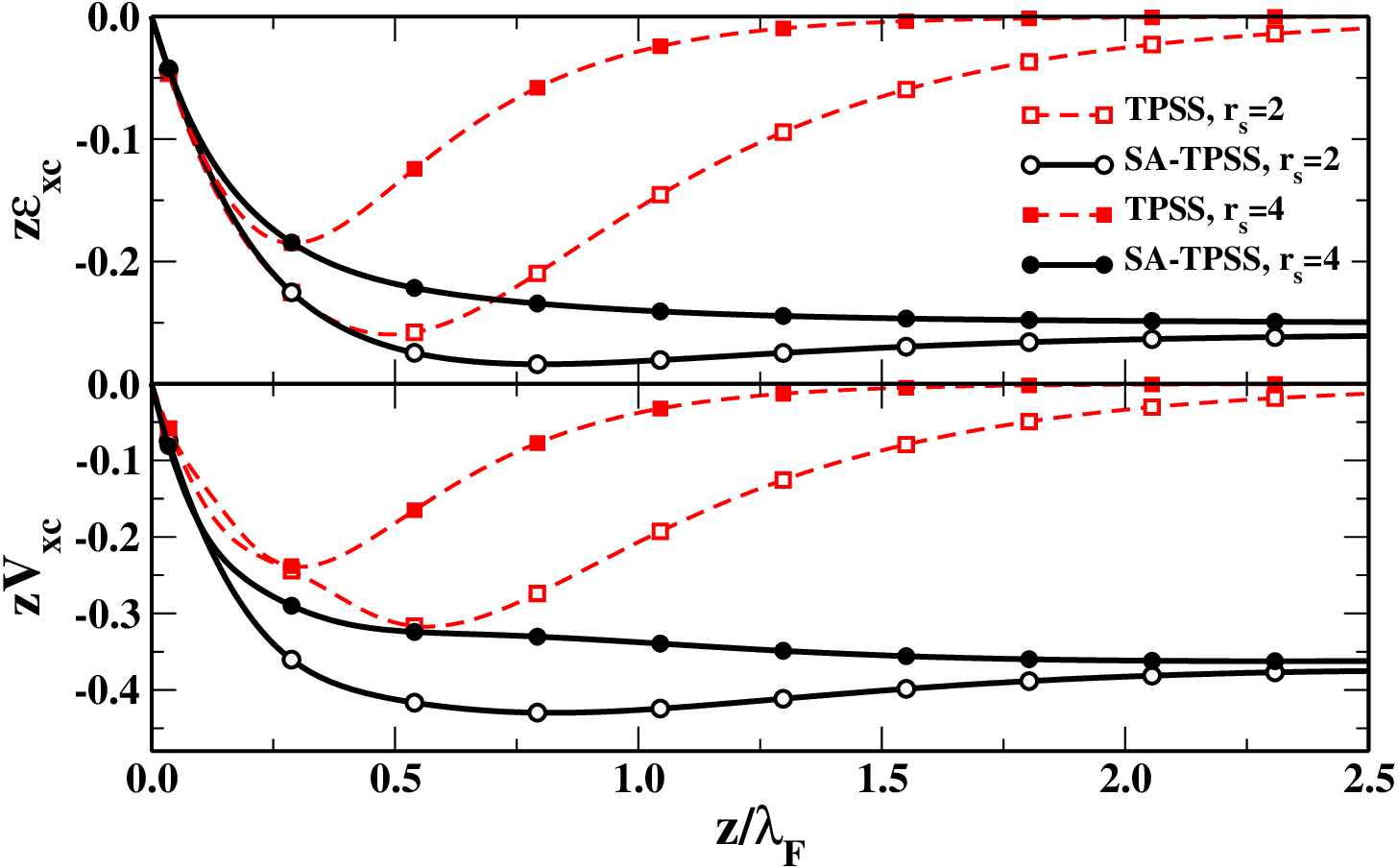}
\caption{$z\epsilon_{xc}(z)$ (upper panel) and $z v_{xc}(z)$ (lower panel) versus the reduced distance 
$z/\lambda_F$ ($\lambda_F$ is the Fermi wavelength), for a jellium surface with electron-density 
parameter $r_s=2$ and $r_s=4$. The surface is located at $z=0$.
Note that SA-TPSS gives $z\epsilon_{xc}=-1/4$ and $z v_{xc}=-3/8$ far outside the surface. 
}
\label{f2}
\end{figure}
%
In Fig. \ref{f2}, fully self-consistent KS-LDA orbitals were used to obtain
$z\epsilon_{xc}(z)$ (upper panel) 
and $z v_{xc}(z)$ (lower panel) as a function of the
scaled distance $z$ for jellium surfaces with the electron-density 
parameters $r_s=2$ and $r_s=4$. 
The TPSS functional yields a wrong (exponential) asymptotic behavior 
for both quantities. 
Instead, the SA-TPSS functional gives the following asymptotic behaviors: 
$\epsilon_{xc}(z)\rightarrow -0.25/z$ and $v_{xc}(z)\rightarrow -0.375/z$. 
Thus, Fig. \ref{f2} provides a numerical proof of the validity of 
Eq. (\ref{em3}), as well as a validation of the simple construction 
used to obtain the SA-TPSS functional.

%
\begin{table}
\begin{center}
\caption{\label{ta1} Jellium surface energies (in erg/cm$^2$), as obtained by using the functionals 
TPPS and SA-TPSS for various values of the electron-density parameter $r_s$.
Diffusion Monte Carlo (DMC) calculations~\cite{WHFGG} are given as a reference.
The results of the popular LDA \cite{kohn1965self,perdewPRB92lda} and PBE \cite{perdewPRL96} functionals, are also shown for 
comparison.
}
\begin{ruledtabular}
\begin{tabular}{cccccc}
$r_s$ & TPSS & SA-TPSS & LDA & PBE &  Ref.  \\
\hline
2 & 3380 & 3368 & 3354 & 3265 & 3392 $\pm$ 50 \\
3 & 772 & 767 & 764 & 741 & 768 $\pm$ 10 \\
4 & 266 & 263 & 261 & 252 & 261 $\pm$ 8 \\
6 & 55.5 & 54.5 & 53 & 52 & 53 $\pm$ ... \\
\end{tabular}
\end{ruledtabular}
\end{center}
\end{table}
%
%
%
\begin{table}
\begin{center}
\caption{\label{ta2} Mean absolute errors (in kcal/mol for energetical tests and in m\AA $\;$ for the bond length test) of several 
molecular properties \cite{bloc,hpbeint}.
}
\begin{ruledtabular}
\begin{tabular}{lcccc}
Test & TPSS & SA-TPSS & LDA & PBE  \\
\hline
atomization energies (W4 test) & 4.7 & 4.8 & 44.0 & 10.7  \\
reaction energies (OMRE test) & 8.0 & 7.9 & 21.0 & 6.7 \\
ionization potentials (IP13 test) & 3.1 & 3.1 & 4.9 & 3.0 \\
bond lengths (MGBL19 test) & 6.9 & 7.0 & 10.0 & 9.3 \\
dipole interactions (DI6 test) & 0.6 & 0.4 & 2.7 & 0.4 \\
hydrogen bonds (HB6 test) & 0.6 & 0.4 & 4.5 & 0.4 \\
\end{tabular}
\end{ruledtabular}
\end{center}
\end{table}
%
In Tables~\ref{ta1} and~\ref{ta2} we report the TPSS and SA-TPSS
results for surface energies and various molecular tests, 
respectively. By construction, 
whenever the surface asymptotics plays a negligible role
(e.g., covalent interactions in molecules: W4, OMRE, IP13, MGBL19) 
both functionals yield very similar results. In the case of
jellium surface energies and non-covalent interactions (DI6, HB6), 
however, SA-TPSS improves over the standard TPSS.

Finally, we have to stress that the SA-TPSS meta-GGA gives only the correct 
asymptotic decay of the XC energy per particle and potential at metal surfaces, but it can not provide
the exact behavior for the exchange and correlation components, separately. This is a difficult task, that in our opinion, a simple 
semilocal functional can not obey.

\section{Airy gas asymptotic properties}

As an additional example of the use of Eq.~(\ref{em3}) and of
the SA-TPSS functional, we consider the Airy gas \cite{KM,AM05}, 
which is the simplest possible model for an edge electron gas (for details see Appendix A).
This model system plays an important role in DFT \cite{AM05,airy3,airy4}, as
it incorporates the correct physics of a semi-infinite metal
surface and is simple enough to allow for analytical calculations.
To our knowledge, the exact asymptotic behavior of the XC energy per particle 
and of KS XC potential of the Airy gas are unknown. Nevertheless, 
we can use the SA-TPSS functional to obtain some information about them.

The Airy-gas electron density and positive-defined kinetic-energy density
are \cite{KM,AM05,airy3,airy2}
\begin{eqnarray}
\label{ae1}
\rho(z) & = &\frac{1}{3\pi}[z^2\rm{Ai}^2(z)-z\rm{Ai'}^2(z)-\rm{Ai}(z)\rm{Ai'}(z)/2], \\
\label{ae2}
\tau & = &-\frac{3}{10}z \rho(z)+\frac{1}{5}\rho''(z) \ ,
\end{eqnarray}
where $z$ is the scaled distance and ${\rm Ai}(z)$ is the Airy function.
%
%
\begin{figure}[t]
\includegraphics[width=\columnwidth]{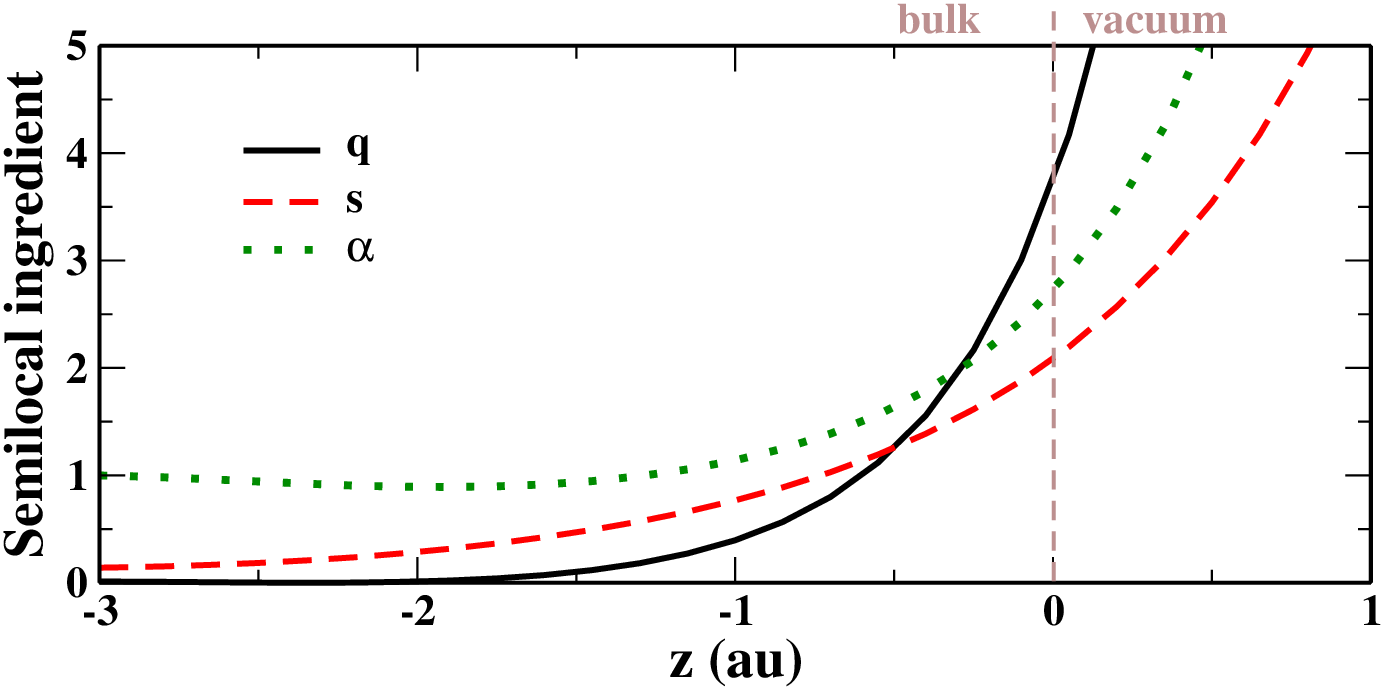}
\caption{The Airy-gas semilocal ingredients ($s$, $q$, and $\alpha$), as a function of the scaled distance $z$.
The Airy bulk is at $z\leq 0$. The exponential decay of the electron density occurs at $z\geq 0$.}
\label{f3}
\end{figure}
%
In Fig.~\ref{f3}, we show the Airy-gas semilocal ingredient $s$, the reduced Laplacian $q=\nabla^2 
\rho/[4(3\pi^{2})^{2/3}\rho^{5/3}]$,
and $\alpha$. In the bulk
($z\rightarrow -\infty$), both $s$ and $q$ are small, while $\alpha\rightarrow 1$ (showing that
the Thomas-Fermi theory becomes exact). In the vacuum, all semilocal ingredients diverge
(as in the case of the jellium surface).

In the limit $z\to\infty$, the Airy-gas electron density and kinetic-energy density are
\begin{eqnarray}
\rho(z\rightarrow\infty) & \rightarrow & \frac{1}{32}\,{\frac { {{\rm e}^{-4/3\,{z}^{3/2}}} }
{{\pi }^{2}z^{3/2}}}-{\frac {35}{768}}\,{\frac {
 {{\rm e}^{-4/3\,{z}^{3/2}}} } {{\pi }^{2}{z}^{3}}}+\ldots \, \nonumber \\
\tau(z\rightarrow\infty) & \rightarrow & {\frac {1}{64}}\,{\frac { {{\rm e}^{-4/3\,{z}^{3/2}}}
}{\sqrt {z}{\pi }^{2}}}+{\frac {13}{1536}}\,{\frac { {
{\rm e}^{-4/3\,{z}^{3/2}}} }{{\pi }^{2}{z}^{2}}}+\ldots\ .
\label{e3}
\end{eqnarray}
The SA-TPSS XC functional gives the following analytical expressions:
\begin{eqnarray}
\epsilon_{xc}^{SA-TPSS}(z\rightarrow\infty) &\rightarrow& -\sqrt{3}/(8 z)\approx -0.217/z,\nonumber \\ 
v^{SA-TPSS}_{xc}(z\rightarrow\infty) & \rightarrow & -3\sqrt{3}/(16 z)\approx -0.325/z.
\label{eaa3}
\end{eqnarray}

\begin{figure}[t]
\includegraphics[width=\columnwidth]{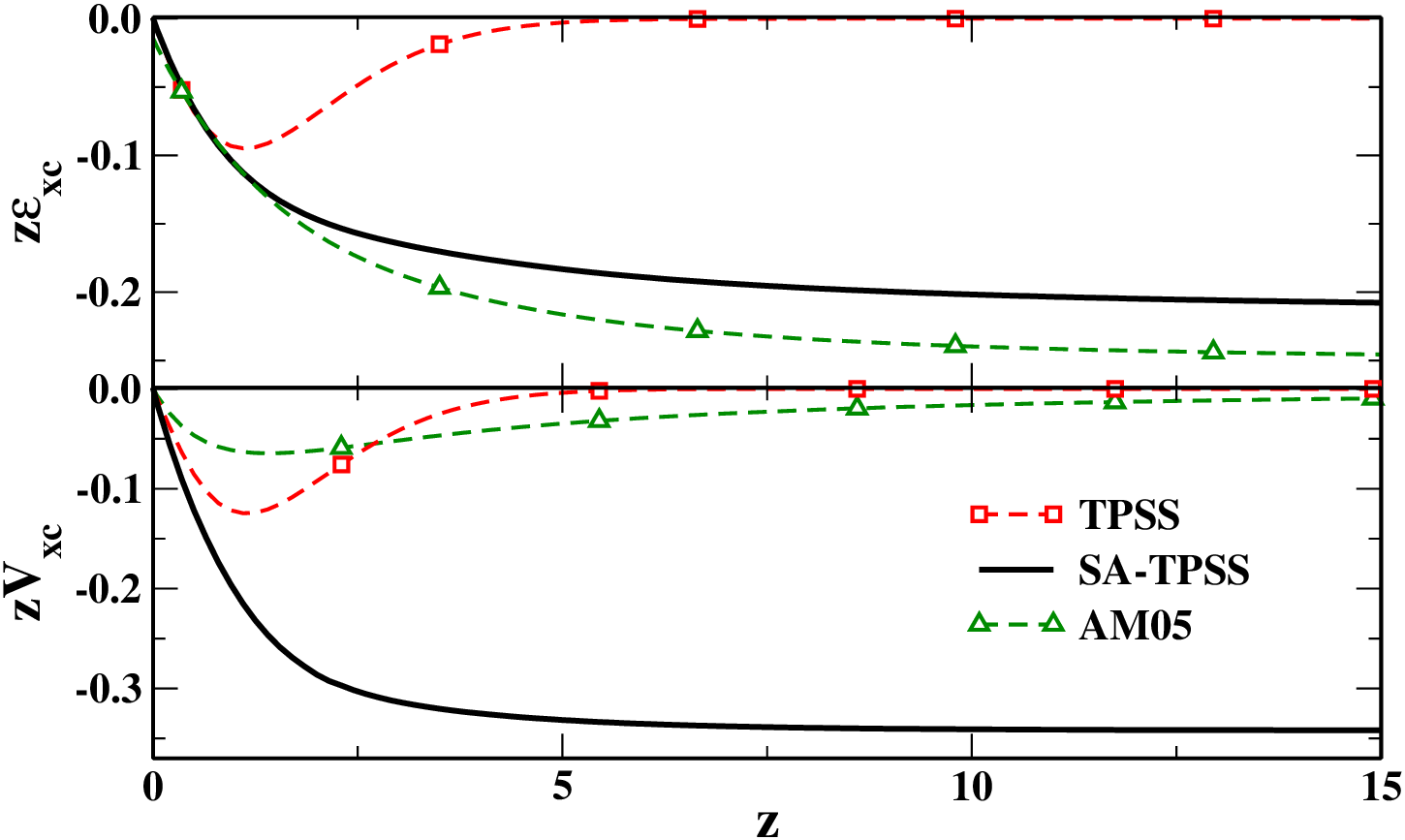}
\caption{The Airy-gas $z\epsilon_{xc}(z)$ (upper panel) and $z v_{xc}(z)$ (lower panel),
as a function of the scaled distance $z$.
}
\label{f4}
\end{figure}

This result is interesting, since it suggests that, for the Airy gas, 
both $\epsilon_{xc}$ and $v_{xc}$ decay as $-1/z$, as in the case of 
the jellium surface.
Furthermore, because the coefficients (0.217 and 0.325) 
are close to, but smaller than, the ones for the 
jellium surface (0.25 and 0.375), 
we can conclude that at a metal surface the main contribution 
to the asymptotics comes from the region near the surface, where the effective 
potential is linear and well described by the Airy-gas model.
We note that such a result cannot be obtained without the
proper inclusion of exact surface conditions into the functional.
This is shown in Fig. \ref{f4} where we plot, for several functionals,
$z\epsilon_{xc}(z)$ (upper panel) and $z v_{xc}(z)$ (lower panel), versus the
scaled distance $z$, in the vacuum region of the Airy gas.
%
%
The TPSS functional shows a rather unphysical exponential decay
for both $\varepsilon_{xc}$ and $v_{xc}$. On the other hand, 
the AM05 functional~\cite{AM05}, whose energy density 
is fitted to the Airy gas, 
is close to our SA-TPSS functional for $\epsilon_{xc}$ but
displays a decay that is too fast for $v_{xc}$. This latter feature
represents a manifestation of the impossibility, at the GGA level,
to describe correctly both $\epsilon_{xc}$ and $v_{xc}$,
which can only be overcome at the meta-GGA level~\cite{alpha}.

The exchange-only asymptotic behaviors ($\epsilon_x$ and $v_x$) at metal surfaces, are depending on 
the bulk and surface parameters ($k_F$ and $W$) \cite{CP3,HPR2,qian2012asymptotic}, and thus they are 
created by bulk- and 
surface-electrons. When correlation is included, screening effects dump the bulk contribution, such that 
the asymptotic properties of $\epsilon_{xc}$ and $v_{xc}$ are mainly created by 
a density-independent XC effect of the surface region, as 
proved by the SA-TPSS result for the Airy gas. Thus, the Airy gas model system can be efficiently used 
in modelling various phenomena outside metal surfaces, even being an alternative to the ad-hoc LDA XC 
potential modifications \cite{serena86,Chulk99}.

\section{Conclusions}

In conclusion, we have derived an exact meta-GGA condition for the 
correct image-like surface
asymptotics of the XC energy per particle $\varepsilon_{xc}$ and the
KS XC potential $v_{xc}$.
Our formula [Eq.~(\ref{em3})] depends only on the semilocal 
ingredient $\alpha$ and
takes advantage of the non-locality of the 
kinetic energy density beyond the von 
Weizs\"{a}cker term \cite{alpha}. The existence of this exact condition
represents an important contribution in the framework of DFT, as it shows that
surface asymptotics can be described by semilocal meta-GGA functionals.
On the contrary, no GGA can be constructed that is able to describe correctly
the asymptotics of both $\varepsilon_{xc}$ and $v_{xc}$. 
In fact, there is, to our knowledge, no GGA functional that yields a 
realistic KS XC potential at metal surfaces.

We have demonstrated that our 
exact condition can be easily 
implemented in any meta-GGA functional, 
keeping its original accuracy for standard ground-state properties
and providing, at the same time, 
a correct description of the surface asymptotics.
We have constructed the SA-TPSS functional, 
which we have shown to perform as the TPSS
for covalent chemistry and to improve 
over it for non-covalent interactions
and surface-related problems. 
This new functional can thus be a promising tool 
for the investigation of surface-sensitive 
electronic-structure calculations, such as 
molecule/molecular complex-surface, cluster-surface, 
and surface-surface interactions.

We thank TURBOMOLE GmbH for providing the TURBOMOLE program package.

  \renewcommand{\theequation}{A-\arabic{equation}}
  \setcounter{equation}{0}  
  \section*{APPENDIX A}  

In case of semi-infinite jellium surfaces, the one-particle eigenfunctions $\phi_{k_z}(z)$
have a continuous
energy spectrum $\epsilon_{k_z}=V_{KS}(\infty)+k_z^2/2$, and are solutions of the one-dimensional
KS equation
\begin{equation}
(-\frac{\partial^2}{2\partial z^2} +V_{KS}(z)-\epsilon_{k_z})\phi_{k_z}(z)=0.
\label{aa1}
\end{equation}
Here $V_{KS}(z)=V_H(z)+V_{xc}(z)$ is the sum of the total classical electrostatic potential
(which incorporates the positive background), and the XC potential.

In case of the Airy gas, the effective potential has the linear form
\begin{equation}
v_{eff}(Z) = \left\{
  \begin{array}{lr}
    -FZ &  \;\rm{when}\;\;\;-\infty< Z < L \\
    +\infty & \;\rm{when}\;\;\; Z \ge L,
  \end{array}
\right.
\label{aa2}
\end{equation} 
with $F>0$ being the slope of the effective potential \cite{KM}. Thus, the one-particle 
normalized eigenfunctions $\phi_j(Z)$ satisfy the one-dimensional equation
\begin{eqnarray}
(-\frac{\partial^2}{2\partial Z^2} -FZ-\epsilon_{j})\phi_{j}(Z)=0,\nonumber\\
\phi_{j}(\infty)=\phi_{j}(L)=0,
\label{aa3}
\end{eqnarray}
being proportional to the Airy function. Here $Z$ is the distance perpendicular to the surface.
It is convenient to consider the scaled distance $z=Z (2F)^{1/3}$ \cite{KM}, as used in the Section IV.


  \renewcommand{\theequation}{B-\arabic{equation}}
  \setcounter{equation}{0}  
  \section*{APPENDIX B}  

\appendix
The electron density and kinetic energy density of a jellim surface are
\begin{eqnarray}
\rho(z) &= &\frac{1}{4\pi^2}\int_{-k_F}^{k_F}dk_z \; (k_F^2-k_z^2)|\phi_{k_z}(z)|^2,
\end{eqnarray}
\begin{eqnarray}
 \tau(z) &= &\int_{-k_F}^{k_F}dk_z \;[\frac{1}{8\pi^2}|\frac{\partial \phi_{k_z}(z)}{\partial z}|^2
(k_F^2-k_z^2)+ \nonumber \\
&& \frac{1}{16\pi^2}|\phi_{k_z}(z)|^2(k_F^2-k_z^2)^2]\ ,
\label{ej2}
\end{eqnarray}
with $k_F$ being the magnitude of the bulk Fermi wave-vector.

Using Eqs. (\ref{ej3}) and (\ref{ej4}) 
we obtain, when $z\rightarrow \infty$, the following expressions 
\begin{widetext}
\begin{equation}
\rho\rightarrow \left( 1/4\,{\frac {{N}^{2}W}{{\it k_F}\,{\pi }^{2}{z}^{2}}}-1/8\,{
\frac {{N}^{2}\sqrt {2}{W}^{3/2}}{{{\it k_F}}^{3}{z}^{3}{\pi }^{2}}}
 \right)  
e^{-2z\sqrt{2W}} z^{2\alpha_{KS}}
%
\label{e1}
\end{equation}

\begin{eqnarray}
&& \tau\rightarrow 1/4\,{\frac {{N}^{2}{W}^{2}}{{\it k_F}\,{\pi }^{2}{z}^{2} 
e^{2z\sqrt{2W}}z^{-2\alpha_{KS}}}}+
 1/32\,{\frac { \left( -8\,{{
\it k_F}}^{4}\sqrt {W}{\it \alpha_{KS}}-4\,{{\it k_F}}^{2}{W}^{3/2}+12\,{{
\it k_F}}^{4}\sqrt {W} \right) {N}^{2}W\,\sqrt {2}}{{{\it k_F}}^{5}{\pi 
}^{2}{z}^{3} 
e^{2z\sqrt{2W}}z^{-2\alpha_{KS}}}}+
\nonumber\\
&& 1/32\,{
\frac { \left( -12\,{{\it k_F}}^{2}W\,\sqrt {2}+3\,\sqrt {2}{{\it k_F}}^
{4}+4\,{{\it k_F}}^{2}W\,\sqrt {2}{\it \alpha_{KS}}-4\,\sqrt {2}{{\it k_F}}^
{4}{\it \alpha_{KS}}+2\,\sqrt {2}{{\it k_F}}^{4}{{\it \alpha_{KS}}}^{2}
 \right) {N}^{2}W\,\sqrt {2}}{{{\it k_F}}^{5}{\pi }^{2}{z}^{4} 
e^{2z\sqrt{2W}}z^{-2\alpha_{KS}}}}+
\nonumber\\
&& 1/32\,{\frac { \left( 6\,{{
\it k_F}}^{2}\sqrt {W}{\it \alpha_{KS}}+6\,{W}^{3/2}-2\,{{\it k_F}}^{2}
\sqrt {W}{{\it \alpha_{KS}}}^{2}-6\,{{\it k_F}}^{2}\sqrt {W} \right) {N}^{2
}W\,\sqrt {2}}{{{\it k_F}}^{5}{\pi }^{2}{z}^{5} 
e^{2z\sqrt{2W}}z^{-2\alpha_{KS}}}}.
\label{e12}
\end{eqnarray}

Then,
\begin{equation}  
\alpha\rightarrow -\,{\frac { \left( -60\,\sqrt [3]{3}\sqrt [6]{2}{{\it
k_F}}^{2}z\,\sqrt {W}+40\,\sqrt [3]{3}{2}^{2/3}W-15\,\sqrt [3]{3}{2}^{2
/3}{{\it k_F}}^{2}
\right) {{\rm e}^{4/3\,z\,\sqrt
{2}\sqrt {W}}}{z}^{-4/3\,{\it \alpha_{KS}}-2/3}{2}^{2/3}}{54 {W}^{2/3}{N}^{4/
3}{{\it k_F}}^{4/3}}},
\label{e2}    
\end{equation}
and
\begin{equation}
F_{xc}\rightarrow 
\frac{4}{9\eta}{\frac { \left( \pi \,{2}^{2/3}{3}^{2/3}\sqrt [3]{{\it k_F}}+9\,O
 \left( {z}^{-4/3} \right) {N}^{2/3}\sqrt [3]{W}\sqrt [3]{z} \right) {
{\rm e}^{2/3\,z\,\sqrt {2}\sqrt {W}}}{z}^{-2/3\,{\it \alpha_{KS}}-1/3}}{{N
}^{2/3}\sqrt [3]{W}}},
\label{e33}
\end{equation}
\end{widetext}
where $N$ is a normalization constant, $W$ is the work function, and $k_F$ is the 
bulk Fermi wave-vector. Here $F_{xc}=\epsilon_{xc}^{MGGA}/\epsilon_{x}^{LDA}$ is the enhancement factor 
corresponding to the energy density
of Eq. (\ref{em3}).

\twocolumngrid
\bibliography{asymeta}

\begin{thebibliography}{55}
\expandafter\ifx\csname natexlab\endcsname\relax\def\natexlab#1{#1}\fi
\expandafter\ifx\csname bibnamefont\endcsname\relax
  \def\bibnamefont#1{#1}\fi
\expandafter\ifx\csname bibfnamefont\endcsname\relax
  \def\bibfnamefont#1{#1}\fi
\expandafter\ifx\csname citenamefont\endcsname\relax
  \def\citenamefont#1{#1}\fi
\expandafter\ifx\csname url\endcsname\relax
  \def\url#1{\texttt{#1}}\fi
\expandafter\ifx\csname urlprefix\endcsname\relax\def\urlprefix{URL }\fi
\providecommand{\bibinfo}[2]{#2}
\providecommand{\eprint}[2][]{\url{#2}}

\bibitem[{\citenamefont{Garcia et~al.}(1985)\citenamefont{Garcia, Reihl, Frank,
  and Williams}}]{garcia85}
\bibinfo{author}{\bibfnamefont{N.}~\bibnamefont{Garcia}},
  \bibinfo{author}{\bibfnamefont{B.}~\bibnamefont{Reihl}},
  \bibinfo{author}{\bibfnamefont{K.~H.} \bibnamefont{Frank}}, \bibnamefont{and}
  \bibinfo{author}{\bibfnamefont{A.~R.} \bibnamefont{Williams}},
  \bibinfo{journal}{Phys. Rev. Lett.} \textbf{\bibinfo{volume}{54}},
  \bibinfo{pages}{591} (\bibinfo{year}{1985}),
  \urlprefix\url{http://link.aps.org/doi/10.1103/PhysRevLett.54.591}.

\bibitem[{\citenamefont{Datta}(2005)}]{datta2005quantum}
\bibinfo{author}{\bibfnamefont{S.}~\bibnamefont{Datta}},
  \emph{\bibinfo{title}{Quantum transport: atom to transistor}}
  (\bibinfo{publisher}{Cambridge University Press}, \bibinfo{year}{2005}).

\bibitem[{\citenamefont{Rundgren and Malmstr\"om}(1977)}]{RM}
\bibinfo{author}{\bibfnamefont{J.}~\bibnamefont{Rundgren}} \bibnamefont{and}
  \bibinfo{author}{\bibfnamefont{G.}~\bibnamefont{Malmstr\"om}},
  \bibinfo{journal}{Phys. Rev. Lett.} \textbf{\bibinfo{volume}{38}},
  \bibinfo{pages}{836} (\bibinfo{year}{1977}).

\bibitem[{\citenamefont{Binnig et~al.}(1985)\citenamefont{Binnig, Frank, Fuchs,
  Garcia, Reihl, Rohrer, Salvan, and Williams}}]{Binning85}
\bibinfo{author}{\bibfnamefont{G.}~\bibnamefont{Binnig}},
  \bibinfo{author}{\bibfnamefont{K.~H.} \bibnamefont{Frank}},
  \bibinfo{author}{\bibfnamefont{H.}~\bibnamefont{Fuchs}},
  \bibinfo{author}{\bibfnamefont{N.}~\bibnamefont{Garcia}},
  \bibinfo{author}{\bibfnamefont{B.}~\bibnamefont{Reihl}},
  \bibinfo{author}{\bibfnamefont{H.}~\bibnamefont{Rohrer}},
  \bibinfo{author}{\bibfnamefont{F.}~\bibnamefont{Salvan}}, \bibnamefont{and}
  \bibinfo{author}{\bibfnamefont{A.~R.} \bibnamefont{Williams}},
  \bibinfo{journal}{Phys. Rev. Lett.} \textbf{\bibinfo{volume}{55}},
  \bibinfo{pages}{991} (\bibinfo{year}{1985}).

\bibitem[{\citenamefont{Pitarke et~al.}(1990)\citenamefont{Pitarke, Flores, and
  Echenique}}]{Pitarke90}
\bibinfo{author}{\bibfnamefont{J.}~\bibnamefont{Pitarke}},
  \bibinfo{author}{\bibfnamefont{F.}~\bibnamefont{Flores}}, \bibnamefont{and}
  \bibinfo{author}{\bibfnamefont{P.}~\bibnamefont{Echenique}},
  \bibinfo{journal}{Surf. Sci.} \textbf{\bibinfo{volume}{234}},
  \bibinfo{pages}{1 } (\bibinfo{year}{1990}), ISSN \bibinfo{issn}{0039-6028}.

\bibitem[{\citenamefont{Harris et~al.}(1997)\citenamefont{Harris, Ge, Jr.,
  McNeill, and Wong}}]{harris97}
\bibinfo{author}{\bibfnamefont{C.~B.} \bibnamefont{Harris}},
  \bibinfo{author}{\bibfnamefont{N.-H.} \bibnamefont{Ge}},
  \bibinfo{author}{\bibfnamefont{R.~L.~L.} \bibnamefont{Jr.}},
  \bibinfo{author}{\bibfnamefont{J.~D.} \bibnamefont{McNeill}},
  \bibnamefont{and} \bibinfo{author}{\bibfnamefont{C.~M.} \bibnamefont{Wong}},
  \bibinfo{journal}{Ann. Rev. Phys. Chem.} \textbf{\bibinfo{volume}{48}},
  \bibinfo{pages}{711} (\bibinfo{year}{1997}).

\bibitem[{\citenamefont{Fauster and Steinmann}(1995)}]{fauster}
\bibinfo{author}{\bibfnamefont{T.}~\bibnamefont{Fauster}} \bibnamefont{and}
  \bibinfo{author}{\bibfnamefont{W.}~\bibnamefont{Steinmann}}, in
  \emph{\bibinfo{booktitle}{Electromagnetic Waves: Recent Development in
  Research, Vol. 2}}, edited by
  \bibinfo{editor}{\bibfnamefont{P.}~\bibnamefont{Halevi}}
  (\bibinfo{publisher}{Elsevier, Amsterdam}, \bibinfo{year}{1995}), p.
  \bibinfo{pages}{350}.

\bibitem[{\citenamefont{Lang and Kohn}(1973)}]{langPRB73}
\bibinfo{author}{\bibfnamefont{N.}~\bibnamefont{Lang}} \bibnamefont{and}
  \bibinfo{author}{\bibfnamefont{W.}~\bibnamefont{Kohn}},
  \bibinfo{journal}{Phys. Rev. B} \textbf{\bibinfo{volume}{7}},
  \bibinfo{pages}{3541} (\bibinfo{year}{1973}).

\bibitem[{\citenamefont{Kohn}(1999)}]{kohn99}
\bibinfo{author}{\bibfnamefont{W.}~\bibnamefont{Kohn}}, \bibinfo{journal}{Rev.
  Mod. Phys.} \textbf{\bibinfo{volume}{71}}, \bibinfo{pages}{1253}
  (\bibinfo{year}{1999}).

\bibitem[{\citenamefont{Kohn and Sham}(1965)}]{kohn1965self}
\bibinfo{author}{\bibfnamefont{W.}~\bibnamefont{Kohn}} \bibnamefont{and}
  \bibinfo{author}{\bibfnamefont{L.~J.} \bibnamefont{Sham}},
  \bibinfo{journal}{Phys. Rev.} \textbf{\bibinfo{volume}{140}},
  \bibinfo{pages}{A1133} (\bibinfo{year}{1965}).

\bibitem[{\citenamefont{Gunnarsson et~al.}(1979)\citenamefont{Gunnarsson,
  Jonson, and Lundqvist}}]{gunn79}
\bibinfo{author}{\bibfnamefont{O.}~\bibnamefont{Gunnarsson}},
  \bibinfo{author}{\bibfnamefont{M.}~\bibnamefont{Jonson}}, \bibnamefont{and}
  \bibinfo{author}{\bibfnamefont{B.~I.} \bibnamefont{Lundqvist}},
  \bibinfo{journal}{Phys. Rev. B} \textbf{\bibinfo{volume}{20}},
  \bibinfo{pages}{3136} (\bibinfo{year}{1979}),
  \urlprefix\url{http://link.aps.org/doi/10.1103/PhysRevB.20.3136}.

\bibitem[{\citenamefont{Eguiluz and Hanke}(1989)}]{eghanke89}
\bibinfo{author}{\bibfnamefont{A.~G.} \bibnamefont{Eguiluz}} \bibnamefont{and}
  \bibinfo{author}{\bibfnamefont{W.}~\bibnamefont{Hanke}},
  \bibinfo{journal}{Phys. Rev. B} \textbf{\bibinfo{volume}{39}},
  \bibinfo{pages}{10433} (\bibinfo{year}{1989}).

\bibitem[{\citenamefont{Eguiluz et~al.}(1992)\citenamefont{Eguiluz,
  Heinrichsmeier, Fleszar, and Hanke}}]{egu92}
\bibinfo{author}{\bibfnamefont{A.~G.} \bibnamefont{Eguiluz}},
  \bibinfo{author}{\bibfnamefont{M.}~\bibnamefont{Heinrichsmeier}},
  \bibinfo{author}{\bibfnamefont{A.}~\bibnamefont{Fleszar}}, \bibnamefont{and}
  \bibinfo{author}{\bibfnamefont{W.}~\bibnamefont{Hanke}},
  \bibinfo{journal}{Phys. Rev. Lett.} \textbf{\bibinfo{volume}{68}},
  \bibinfo{pages}{1359} (\bibinfo{year}{1992}).

\bibitem[{\citenamefont{White et~al.}(1998)\citenamefont{White, Godby, Rieger,
  and Needs}}]{white98}
\bibinfo{author}{\bibfnamefont{I.~D.} \bibnamefont{White}},
  \bibinfo{author}{\bibfnamefont{R.~W.} \bibnamefont{Godby}},
  \bibinfo{author}{\bibfnamefont{M.~M.} \bibnamefont{Rieger}},
  \bibnamefont{and} \bibinfo{author}{\bibfnamefont{R.~J.} \bibnamefont{Needs}},
  \bibinfo{journal}{Phys. Rev. Lett.} \textbf{\bibinfo{volume}{80}},
  \bibinfo{pages}{4265} (\bibinfo{year}{1998}).

\bibitem[{\citenamefont{Qian and Sahni}(2005)}]{qian2005exact}
\bibinfo{author}{\bibfnamefont{Z.}~\bibnamefont{Qian}} \bibnamefont{and}
  \bibinfo{author}{\bibfnamefont{V.}~\bibnamefont{Sahni}},
  \bibinfo{journal}{Int. J. Quantum Chem.} \textbf{\bibinfo{volume}{104}},
  \bibinfo{pages}{929} (\bibinfo{year}{2005}).

\bibitem[{\citenamefont{Hoeft et~al.}(2001)\citenamefont{Hoeft, Kittel, Polcik,
  Bao, Toomes, Kang, Woodruff, Pascal, and Lamont}}]{hoeft2001molecular}
\bibinfo{author}{\bibfnamefont{J.-T.} \bibnamefont{Hoeft}},
  \bibinfo{author}{\bibfnamefont{M.}~\bibnamefont{Kittel}},
  \bibinfo{author}{\bibfnamefont{M.}~\bibnamefont{Polcik}},
  \bibinfo{author}{\bibfnamefont{S.}~\bibnamefont{Bao}},
  \bibinfo{author}{\bibfnamefont{R.~L.} \bibnamefont{Toomes}},
  \bibinfo{author}{\bibfnamefont{J.-H.} \bibnamefont{Kang}},
  \bibinfo{author}{\bibfnamefont{D.~P.} \bibnamefont{Woodruff}},
  \bibinfo{author}{\bibfnamefont{M.}~\bibnamefont{Pascal}}, \bibnamefont{and}
  \bibinfo{author}{\bibfnamefont{C.~L.~A.} \bibnamefont{Lamont}},
  \bibinfo{journal}{Phys. Rev. Lett.} \textbf{\bibinfo{volume}{87}},
  \bibinfo{pages}{086101} (\bibinfo{year}{2001}).

\bibitem[{\citenamefont{Becke}(1988)}]{b88}
\bibinfo{author}{\bibfnamefont{A.~D.} \bibnamefont{Becke}},
  \bibinfo{journal}{Phys. Rev. A} \textbf{\bibinfo{volume}{38}},
  \bibinfo{pages}{3098} (\bibinfo{year}{1988}).

\bibitem[{\citenamefont{Serena et~al.}(1986)\citenamefont{Serena, Soler, and
  Garc\'{\i}a}}]{serena86}
\bibinfo{author}{\bibfnamefont{P.~A.} \bibnamefont{Serena}},
  \bibinfo{author}{\bibfnamefont{J.~M.} \bibnamefont{Soler}}, \bibnamefont{and}
  \bibinfo{author}{\bibfnamefont{N.}~\bibnamefont{Garc\'{\i}a}},
  \bibinfo{journal}{Phys. Rev. B} \textbf{\bibinfo{volume}{34}},
  \bibinfo{pages}{6767} (\bibinfo{year}{1986}),
  \urlprefix\url{http://link.aps.org/doi/10.1103/PhysRevB.34.6767}.

\bibitem[{\citenamefont{Chulkov et~al.}(1999)\citenamefont{Chulkov, Silkin, and
  Echenique}}]{Chulk99}
\bibinfo{author}{\bibfnamefont{E.}~\bibnamefont{Chulkov}},
  \bibinfo{author}{\bibfnamefont{V.}~\bibnamefont{Silkin}}, \bibnamefont{and}
  \bibinfo{author}{\bibfnamefont{P.}~\bibnamefont{Echenique}},
  \bibinfo{journal}{Surface Science} \textbf{\bibinfo{volume}{437}},
  \bibinfo{pages}{330 } (\bibinfo{year}{1999}), ISSN \bibinfo{issn}{0039-6028}.

\bibitem[{\citenamefont{Gunnarsson and Jones}(1980)}]{gunn80}
\bibinfo{author}{\bibfnamefont{O.}~\bibnamefont{Gunnarsson}} \bibnamefont{and}
  \bibinfo{author}{\bibfnamefont{R.~O.} \bibnamefont{Jones}},
  \bibinfo{journal}{Physica Scripta} \textbf{\bibinfo{volume}{21}},
  \bibinfo{pages}{394} (\bibinfo{year}{1980}),
  \urlprefix\url{http://iopscience.iop.org/1402-4896/21/3-4/027}.

\bibitem[{\citenamefont{Ossicini et~al.}(1986)\citenamefont{Ossicini, Bertoni,
  and Gies}}]{ossicini86}
\bibinfo{author}{\bibfnamefont{S.}~\bibnamefont{Ossicini}},
  \bibinfo{author}{\bibfnamefont{C.~M.} \bibnamefont{Bertoni}},
  \bibnamefont{and} \bibinfo{author}{\bibfnamefont{P.}~\bibnamefont{Gies}},
  \bibinfo{journal}{Europhy. Lett.} \textbf{\bibinfo{volume}{1}},
  \bibinfo{pages}{661} (\bibinfo{year}{1986}).

\bibitem[{\citenamefont{Garc\'{\i}a-Gonz\'alez
  et~al.}(2000)\citenamefont{Garc\'{\i}a-Gonz\'alez, Alvarellos, Chac\'on, and
  Tarazona}}]{garcia00}
\bibinfo{author}{\bibfnamefont{P.}~\bibnamefont{Garc\'{\i}a-Gonz\'alez}},
  \bibinfo{author}{\bibfnamefont{J.~E.} \bibnamefont{Alvarellos}},
  \bibinfo{author}{\bibfnamefont{E.}~\bibnamefont{Chac\'on}}, \bibnamefont{and}
  \bibinfo{author}{\bibfnamefont{P.}~\bibnamefont{Tarazona}},
  \bibinfo{journal}{Phys. Rev. B} \textbf{\bibinfo{volume}{62}},
  \bibinfo{pages}{16063} (\bibinfo{year}{2000}).

\bibitem[{\citenamefont{Harris and Griffin}(1975)}]{HG}
\bibinfo{author}{\bibfnamefont{J.}~\bibnamefont{Harris}} \bibnamefont{and}
  \bibinfo{author}{\bibfnamefont{A.}~\bibnamefont{Griffin}},
  \bibinfo{journal}{Phys. Rev. B} \textbf{\bibinfo{volume}{11}},
  \bibinfo{pages}{3669} (\bibinfo{year}{1975}).

\bibitem[{\citenamefont{Langreth and Perdew}(1977)}]{LP}
\bibinfo{author}{\bibfnamefont{D.~C.} \bibnamefont{Langreth}} \bibnamefont{and}
  \bibinfo{author}{\bibfnamefont{J.~P.} \bibnamefont{Perdew}},
  \bibinfo{journal}{Phys. Rev. B} \textbf{\bibinfo{volume}{15}},
  \bibinfo{pages}{2884} (\bibinfo{year}{1977}).

\bibitem[{\citenamefont{Gunnarsson and Lundqvist}(1976)}]{GL}
\bibinfo{author}{\bibfnamefont{O.}~\bibnamefont{Gunnarsson}} \bibnamefont{and}
  \bibinfo{author}{\bibfnamefont{B.~I.} \bibnamefont{Lundqvist}},
  \bibinfo{journal}{Phys. Rev. B} \textbf{\bibinfo{volume}{13}},
  \bibinfo{pages}{4274} (\bibinfo{year}{1976}).

\bibitem[{\citenamefont{Constantin and Pitarke}(2011)}]{CP2}
\bibinfo{author}{\bibfnamefont{L.~A.} \bibnamefont{Constantin}}
  \bibnamefont{and} \bibinfo{author}{\bibfnamefont{J.~M.}
  \bibnamefont{Pitarke}}, \bibinfo{journal}{Phys. Rev. B}
  \textbf{\bibinfo{volume}{83}}, \bibinfo{pages}{075116}
  (\bibinfo{year}{2011}).

\bibitem[{\citenamefont{Horowitz et~al.}(2009)\citenamefont{Horowitz,
  Constantin, Proetto, and Pitarke}}]{CP3}
\bibinfo{author}{\bibfnamefont{C.~M.} \bibnamefont{Horowitz}},
  \bibinfo{author}{\bibfnamefont{L.~A.} \bibnamefont{Constantin}},
  \bibinfo{author}{\bibfnamefont{C.~R.} \bibnamefont{Proetto}},
  \bibnamefont{and} \bibinfo{author}{\bibfnamefont{J.~M.}
  \bibnamefont{Pitarke}}, \bibinfo{journal}{Phys. Rev. B}
  \textbf{\bibinfo{volume}{80}}, \bibinfo{pages}{235101}
  (\bibinfo{year}{2009}).

\bibitem[{\citenamefont{Qian}(2012)}]{qian2012asymptotic}
\bibinfo{author}{\bibfnamefont{Z.}~\bibnamefont{Qian}}, \bibinfo{journal}{Phys.
  Rev. B} \textbf{\bibinfo{volume}{85}}, \bibinfo{pages}{115124}
  (\bibinfo{year}{2012}).

\bibitem[{\citenamefont{Horowitz et~al.}(2010)\citenamefont{Horowitz, Proetto,
  and Pitarke}}]{HPR2}
\bibinfo{author}{\bibfnamefont{C.~M.} \bibnamefont{Horowitz}},
  \bibinfo{author}{\bibfnamefont{C.~R.} \bibnamefont{Proetto}},
  \bibnamefont{and} \bibinfo{author}{\bibfnamefont{J.~M.}
  \bibnamefont{Pitarke}}, \bibinfo{journal}{Phys. Rev. B}
  \textbf{\bibinfo{volume}{81}}, \bibinfo{pages}{121106}
  (\bibinfo{year}{2010}).

\bibitem[{\citenamefont{Horowitz et~al.}(2006)\citenamefont{Horowitz, Proetto,
  and Rigamonti}}]{horo06}
\bibinfo{author}{\bibfnamefont{C.~M.} \bibnamefont{Horowitz}},
  \bibinfo{author}{\bibfnamefont{C.~R.} \bibnamefont{Proetto}},
  \bibnamefont{and}
  \bibinfo{author}{\bibfnamefont{S.}~\bibnamefont{Rigamonti}},
  \bibinfo{journal}{Phys. Rev. Lett.} \textbf{\bibinfo{volume}{97}},
  \bibinfo{pages}{026802} (\bibinfo{year}{2006}),
  \urlprefix\url{http://link.aps.org/doi/10.1103/PhysRevLett.97.026802}.

\bibitem[{\citenamefont{Horowitz et~al.}(2008)\citenamefont{Horowitz, Proetto,
  and Pitarke}}]{horo08}
\bibinfo{author}{\bibfnamefont{C.~M.} \bibnamefont{Horowitz}},
  \bibinfo{author}{\bibfnamefont{C.~R.} \bibnamefont{Proetto}},
  \bibnamefont{and} \bibinfo{author}{\bibfnamefont{J.~M.}
  \bibnamefont{Pitarke}}, \bibinfo{journal}{Phys. Rev. B}
  \textbf{\bibinfo{volume}{78}}, \bibinfo{pages}{085126}
  (\bibinfo{year}{2008}),
  \urlprefix\url{http://link.aps.org/doi/10.1103/PhysRevB.78.085126}.

\bibitem[{\citenamefont{Ye}(2015)}]{ye15}
\bibinfo{author}{\bibfnamefont{L.-H.} \bibnamefont{Ye}},
  \bibinfo{journal}{Phys. Rev. B} \textbf{\bibinfo{volume}{92}},
  \bibinfo{pages}{115132} (\bibinfo{year}{2015}).

\bibitem[{\citenamefont{Engel}(2014{\natexlab{a}})}]{engel2014exact}
\bibinfo{author}{\bibfnamefont{E.}~\bibnamefont{Engel}}, \bibinfo{journal}{J.
  Chem. Phys.} \textbf{\bibinfo{volume}{140}}, \bibinfo{pages}{18A505}
  (\bibinfo{year}{2014}{\natexlab{a}}).

\bibitem[{\citenamefont{Engel}(2014{\natexlab{b}})}]{engel2014asymptotic}
\bibinfo{author}{\bibfnamefont{E.}~\bibnamefont{Engel}},
  \bibinfo{journal}{Phys. Rev. B} \textbf{\bibinfo{volume}{89}},
  \bibinfo{pages}{245105} (\bibinfo{year}{2014}{\natexlab{b}}).

\bibitem[{\citenamefont{Engel et~al.}(1992)\citenamefont{Engel, Chevary,
  Macdonald, and Vosko}}]{engel92}
\bibinfo{author}{\bibfnamefont{E.}~\bibnamefont{Engel}},
  \bibinfo{author}{\bibfnamefont{J.}~\bibnamefont{Chevary}},
  \bibinfo{author}{\bibfnamefont{L.}~\bibnamefont{Macdonald}},
  \bibnamefont{and} \bibinfo{author}{\bibfnamefont{S.}~\bibnamefont{Vosko}},
  \bibinfo{journal}{Z. Phys. D} \textbf{\bibinfo{volume}{23}},
  \bibinfo{pages}{7} (\bibinfo{year}{1992}), ISSN \bibinfo{issn}{0178-7683}.

\bibitem[{\citenamefont{van Leeuwen and Baerends}(1994)}]{lb94}
\bibinfo{author}{\bibfnamefont{R.}~\bibnamefont{van Leeuwen}} \bibnamefont{and}
  \bibinfo{author}{\bibfnamefont{E.~J.} \bibnamefont{Baerends}},
  \bibinfo{journal}{Phys. Rev. A} \textbf{\bibinfo{volume}{49}},
  \bibinfo{pages}{2421} (\bibinfo{year}{1994}).

\bibitem[{\citenamefont{Armiento and K\"ummel}(2013)}]{arm13}
\bibinfo{author}{\bibfnamefont{R.}~\bibnamefont{Armiento}} \bibnamefont{and}
  \bibinfo{author}{\bibfnamefont{S.}~\bibnamefont{K\"ummel}},
  \bibinfo{journal}{Phys. Rev. Lett.} \textbf{\bibinfo{volume}{111}},
  \bibinfo{pages}{036402} (\bibinfo{year}{2013}).

\bibitem[{\citenamefont{Della~Sala et~al.}(2015)\citenamefont{Della~Sala,
  Fabiano, and Constantin}}]{alpha}
\bibinfo{author}{\bibfnamefont{F.}~\bibnamefont{Della~Sala}},
  \bibinfo{author}{\bibfnamefont{E.}~\bibnamefont{Fabiano}}, \bibnamefont{and}
  \bibinfo{author}{\bibfnamefont{L.~A.} \bibnamefont{Constantin}},
  \bibinfo{journal}{Phys. Rev. B} \textbf{\bibinfo{volume}{91}},
  \bibinfo{pages}{035126} (\bibinfo{year}{2015}).

\bibitem[{\citenamefont{Lang and Kohn}(1970)}]{LK}
\bibinfo{author}{\bibfnamefont{N.~D.} \bibnamefont{Lang}} \bibnamefont{and}
  \bibinfo{author}{\bibfnamefont{W.}~\bibnamefont{Kohn}},
  \bibinfo{journal}{Phys. Rev. B} \textbf{\bibinfo{volume}{1}},
  \bibinfo{pages}{4555} (\bibinfo{year}{1970}).

\bibitem[{\citenamefont{Lang and Kohn}(1971)}]{langPRB71}
\bibinfo{author}{\bibfnamefont{N.}~\bibnamefont{Lang}} \bibnamefont{and}
  \bibinfo{author}{\bibfnamefont{W.}~\bibnamefont{Kohn}},
  \bibinfo{journal}{Phys. Rev. B} \textbf{\bibinfo{volume}{3}},
  \bibinfo{pages}{1215} (\bibinfo{year}{1971}).

\bibitem[{\citenamefont{Silvi and Savin}(1994)}]{elf}
\bibinfo{author}{\bibfnamefont{B.}~\bibnamefont{Silvi}} \bibnamefont{and}
  \bibinfo{author}{\bibfnamefont{A.}~\bibnamefont{Savin}},
  \bibinfo{journal}{Nature} \textbf{\bibinfo{volume}{371}},
  \bibinfo{pages}{683} (\bibinfo{year}{1994}).

\bibitem[{\citenamefont{Arbuznikov et~al.}(2002)\citenamefont{Arbuznikov,
  Kaupp, Malkin, Reviakine, and Malkina}}]{arb02}
\bibinfo{author}{\bibfnamefont{A.~V.} \bibnamefont{Arbuznikov}},
  \bibinfo{author}{\bibfnamefont{M.}~\bibnamefont{Kaupp}},
  \bibinfo{author}{\bibfnamefont{V.~G.} \bibnamefont{Malkin}},
  \bibinfo{author}{\bibfnamefont{R.}~\bibnamefont{Reviakine}},
  \bibnamefont{and} \bibinfo{author}{\bibfnamefont{O.~L.}
  \bibnamefont{Malkina}}, \bibinfo{journal}{Phys. Chem. Chem. Phys.}
  \textbf{\bibinfo{volume}{4}}, \bibinfo{pages}{5467} (\bibinfo{year}{2002}).

\bibitem[{\citenamefont{Tao et~al.}(2003)\citenamefont{Tao, Perdew, Staroverov,
  and Scuseria}}]{tpss}
\bibinfo{author}{\bibfnamefont{J.}~\bibnamefont{Tao}},
  \bibinfo{author}{\bibfnamefont{J.~P.} \bibnamefont{Perdew}},
  \bibinfo{author}{\bibfnamefont{V.~N.} \bibnamefont{Staroverov}},
  \bibnamefont{and} \bibinfo{author}{\bibfnamefont{G.~E.}
  \bibnamefont{Scuseria}}, \bibinfo{journal}{Phys. Rev. Lett.}
  \textbf{\bibinfo{volume}{91}}, \bibinfo{pages}{146401}
  (\bibinfo{year}{2003}).

\bibitem[{\citenamefont{Constantin et~al.}(2015)\citenamefont{Constantin,
  Terentjevs, Della~Sala, and Fabiano}}]{sll}
\bibinfo{author}{\bibfnamefont{L.~A.} \bibnamefont{Constantin}},
  \bibinfo{author}{\bibfnamefont{A.}~\bibnamefont{Terentjevs}},
  \bibinfo{author}{\bibfnamefont{F.}~\bibnamefont{Della~Sala}},
  \bibnamefont{and} \bibinfo{author}{\bibfnamefont{E.}~\bibnamefont{Fabiano}},
  \bibinfo{journal}{Phys. Rev. B} \textbf{\bibinfo{volume}{91}},
  \bibinfo{pages}{041120} (\bibinfo{year}{2015}).

\bibitem[{\citenamefont{Feinblum et~al.}(2014)\citenamefont{Feinblum, Kenison,
  and Burke}}]{feinblum2014communication}
\bibinfo{author}{\bibfnamefont{D.~V.} \bibnamefont{Feinblum}},
  \bibinfo{author}{\bibfnamefont{J.}~\bibnamefont{Kenison}}, \bibnamefont{and}
  \bibinfo{author}{\bibfnamefont{K.}~\bibnamefont{Burke}}, \bibinfo{journal}{J.
  Chem. Phys.} \textbf{\bibinfo{volume}{141}}, \bibinfo{pages}{241105}
  (\bibinfo{year}{2014}).

\bibitem[{\citenamefont{Wood et~al.}(2007)\citenamefont{Wood, Hine, Foulkes,
  and Garc\'{\i}a-Gonz\'alez}}]{WHFGG}
\bibinfo{author}{\bibfnamefont{B.}~\bibnamefont{Wood}},
  \bibinfo{author}{\bibfnamefont{N.~D.~M.} \bibnamefont{Hine}},
  \bibinfo{author}{\bibfnamefont{W.~M.~C.} \bibnamefont{Foulkes}},
  \bibnamefont{and}
  \bibinfo{author}{\bibfnamefont{P.}~\bibnamefont{Garc\'{\i}a-Gonz\'alez}},
  \bibinfo{journal}{Phys. Rev. B} \textbf{\bibinfo{volume}{76}},
  \bibinfo{pages}{035403} (\bibinfo{year}{2007}).

\bibitem[{\citenamefont{Perdew and Wang}(1992)}]{perdewPRB92lda}
\bibinfo{author}{\bibfnamefont{J.~P.} \bibnamefont{Perdew}} \bibnamefont{and}
  \bibinfo{author}{\bibfnamefont{Y.}~\bibnamefont{Wang}},
  \bibinfo{journal}{Phys. Rev. B} \textbf{\bibinfo{volume}{45}},
  \bibinfo{pages}{13244} (\bibinfo{year}{1992}).

\bibitem[{\citenamefont{Perdew et~al.}(1996)\citenamefont{Perdew, Burke, and
  Ernzerhof}}]{perdewPRL96}
\bibinfo{author}{\bibfnamefont{J.~P.} \bibnamefont{Perdew}},
  \bibinfo{author}{\bibfnamefont{K.}~\bibnamefont{Burke}}, \bibnamefont{and}
  \bibinfo{author}{\bibfnamefont{M.}~\bibnamefont{Ernzerhof}},
  \bibinfo{journal}{Phys. Rev. Lett.} \textbf{\bibinfo{volume}{77}},
  \bibinfo{pages}{3865} (\bibinfo{year}{1996}).

\bibitem[{\citenamefont{Constantin et~al.}(2013)\citenamefont{Constantin,
  Fabiano, and Sala}}]{bloc}
\bibinfo{author}{\bibfnamefont{L.~A.} \bibnamefont{Constantin}},
  \bibinfo{author}{\bibfnamefont{E.}~\bibnamefont{Fabiano}}, \bibnamefont{and}
  \bibinfo{author}{\bibfnamefont{F.~D.} \bibnamefont{Sala}},
  \bibinfo{journal}{J. Chem. Theory Comput.} \textbf{\bibinfo{volume}{9}},
  \bibinfo{pages}{2256} (\bibinfo{year}{2013}).

\bibitem[{\citenamefont{Fabiano et~al.}(2013)\citenamefont{Fabiano, Constantin,
  and Della~Sala}}]{hpbeint}
\bibinfo{author}{\bibfnamefont{E.}~\bibnamefont{Fabiano}},
  \bibinfo{author}{\bibfnamefont{L.~A.} \bibnamefont{Constantin}},
  \bibnamefont{and}
  \bibinfo{author}{\bibfnamefont{F.}~\bibnamefont{Della~Sala}},
  \bibinfo{journal}{Int. J. Quantum Chem.} \textbf{\bibinfo{volume}{113}},
  \bibinfo{pages}{673} (\bibinfo{year}{2013}), ISSN \bibinfo{issn}{1097-461X},
  \urlprefix\url{http://dx.doi.org/10.1002/qua.24042}.

\bibitem[{\citenamefont{Kohn and Mattsson}(1998)}]{KM}
\bibinfo{author}{\bibfnamefont{W.}~\bibnamefont{Kohn}} \bibnamefont{and}
  \bibinfo{author}{\bibfnamefont{A.~E.} \bibnamefont{Mattsson}},
  \bibinfo{journal}{Phys. Rev. Lett.} \textbf{\bibinfo{volume}{81}},
  \bibinfo{pages}{3487} (\bibinfo{year}{1998}).

\bibitem[{\citenamefont{Armiento and Mattsson}(2005)}]{AM05}
\bibinfo{author}{\bibfnamefont{R.}~\bibnamefont{Armiento}} \bibnamefont{and}
  \bibinfo{author}{\bibfnamefont{A.~E.} \bibnamefont{Mattsson}},
  \bibinfo{journal}{Phys. Rev. B} \textbf{\bibinfo{volume}{72}},
  \bibinfo{pages}{085108} (\bibinfo{year}{2005}).

\bibitem[{\citenamefont{Vitos et~al.}(2000{\natexlab{a}})\citenamefont{Vitos,
  Johansson, Koll\'ar, and Skriver}}]{airy3}
\bibinfo{author}{\bibfnamefont{L.}~\bibnamefont{Vitos}},
  \bibinfo{author}{\bibfnamefont{B.}~\bibnamefont{Johansson}},
  \bibinfo{author}{\bibfnamefont{J.}~\bibnamefont{Koll\'ar}}, \bibnamefont{and}
  \bibinfo{author}{\bibfnamefont{H.~L.} \bibnamefont{Skriver}},
  \bibinfo{journal}{Phys. Rev. A} \textbf{\bibinfo{volume}{61}},
  \bibinfo{pages}{052511} (\bibinfo{year}{2000}{\natexlab{a}}).

\bibitem[{\citenamefont{Constantin and Ruzsinszky}(2009)}]{airy4}
\bibinfo{author}{\bibfnamefont{L.~A.} \bibnamefont{Constantin}}
  \bibnamefont{and}
  \bibinfo{author}{\bibfnamefont{A.}~\bibnamefont{Ruzsinszky}},
  \bibinfo{journal}{Phys. Rev. B} \textbf{\bibinfo{volume}{79}},
  \bibinfo{pages}{115117} (\bibinfo{year}{2009}).

\bibitem[{\citenamefont{Vitos et~al.}(2000{\natexlab{b}})\citenamefont{Vitos,
  Johansson, Koll\'ar, and Skriver}}]{airy2}
\bibinfo{author}{\bibfnamefont{L.}~\bibnamefont{Vitos}},
  \bibinfo{author}{\bibfnamefont{B.}~\bibnamefont{Johansson}},
  \bibinfo{author}{\bibfnamefont{J.}~\bibnamefont{Koll\'ar}}, \bibnamefont{and}
  \bibinfo{author}{\bibfnamefont{H.~L.} \bibnamefont{Skriver}},
  \bibinfo{journal}{Phys. Rev. B} \textbf{\bibinfo{volume}{62}},
  \bibinfo{pages}{10046} (\bibinfo{year}{2000}{\natexlab{b}}).

\end{thebibliography}
\end{document}